\documentclass[aps,prb,superscriptaddress,showpacs,twocolumn,citeautoscript,notitlepage,longbibliography]{revtex4-1}

\usepackage{listings}

\usepackage{graphicx}
\usepackage{amsmath}
\usepackage{amssymb}
\usepackage[normalem]{ulem}
\usepackage{braket}
\usepackage{color}
\usepackage{gensymb}
\usepackage[normalem]{ulem}

\usepackage{makecell}
\usepackage{dsfont}
\usepackage{mathtools}
\newcommand{\bea}{\begin{eqnarray*}}
\newcommand{\eea}{\end{eqnarray*}}
\newcommand{\bne}{\begin{equation*}}
\newcommand{\ede}{\end{equation*}}

\newcommand{\bnen}{\begin{equation}}
\newcommand{\eden}{\end{equation}}
\newcommand{\bean}{\begin{eqnarray}}
\newcommand{\eean}{\end{eqnarray}}
\newcommand{\bsen}{\begin{subequations}}
\newcommand{\esen}{\end{subequations}}

\newcommand{\bna}{\begin{array}}
\newcommand{\eda}{\end{array}}
\newcommand{\bnm}{\begin{enumerate}}
\newcommand{\edm}{\end{enumerate}}
\newcommand{\bni}{\begin{itemize}}
\newcommand{\edi}{\end{itemize}}

\renewcommand{\vec}[1]{\text{\boldmath{$ #1 $}}}

\begin{document}

\title{Magnetic degeneracy points in interacting two-spin systems: \\
geometrical patterns, topological charge distributions, and their stability}

\author{Gy\"orgy~Frank}
\affiliation{Department of Physics, Budapest University of Technology and Economics and MTA-BME "Momentum" Nanoelectronics Research Group, H-1111 Budapest, Budafoki \'ut 8., Hungary}

\author{Zolt\'an~Scher\"ubl}
\affiliation{Department of Physics, Budapest University of Technology and Economics and MTA-BME "Momentum" Nanoelectronics Research Group, H-1111 Budapest, Budafoki \'ut 8., Hungary}

\author{Szabolcs~Csonka}
\affiliation{Department of Physics, Budapest University of Technology and Economics and MTA-BME "Momentum" Nanoelectronics Research Group, H-1111 Budapest, Budafoki \'ut 8., Hungary}

\author{Gergely~Zar\'and}
\affiliation{Department of Theoretical Physics and
MTA-BME Exotic Quantum Phases "Momentum" Research Group, Budapest University of Technology and Economics, 
H-1111 Budapest, Hungary}
\affiliation{
MTA-BME Quantum Correlations Group, Budapest University of Technology and Economics,  H-1111 Budapest, Hungary}

\author{Andr\'as~P\'alyi}
\email{palyi@mail.bme.hu}
\affiliation{Department of Theoretical Physics and
MTA-BME Exotic Quantum Phases "Momentum" Research Group, Budapest University of Technology and Economics, 
H-1111 Budapest, Hungary}

\date{\today}

\begin{abstract}
Spectral degeneracies of quantum magnets are often described
as diabolical points or magnetic Weyl points, which carry
 topological charge. 
Here, we study a simple, yet experimentally 
relevant quantum magnet:
two localized interacting electrons subject to spin-orbit coupling.
In this setting, the degeneracies are not necessarily isolated points,
but can also form a line or a surface. 
We identify ten different possible geometrical patterns
formed by these degeneracy points, 
and study their stability under small perturbations
of the Hamiltonian. 
Stable structures are found to depend on the relative sign of the determinants of the two $g$-tensors, $\cal S$.
Both for ${\cal S}=+1$ and ${\cal S}=-1$, two stable configurations 
are found, and three out of these four configurations are 
formed by pairs of Weyl points. 
These stable 
regions are separated by a surface 
of almost stable configurations, with a structure akin to co-dimension one 
bifurcations.
\end{abstract}

\maketitle


\section{Introduction.}
Nuclear and electron spins are ubiquitous constituents 
in condensed matter physics.
Already a few interacting quantum spins exhibit
a rich variety of phenomena in rather different settings such as molecular 
magnets \cite{Wernsdorfer,Garg,Bruno}, 
magnetic adatoms \cite{Wiesendanger,Spinelli}, 
or spin-based quantum bits \cite{Hanson,Zwanenburg}, to name a few.
When studied in the three-dimensional parameter space 
defined by an external magnetic field, 
all these quantum magnets
possess an intrinsic geometrical 
and topological structure, characterized by concepts \cite{Berry,Wilczek}
such as the Berry phase, the Berry curvature, and the Chern number.
This geometrical structure plays an important role in  
coherent dynamics \cite{SanJosePRB} as well as in
decoherence effects \cite{SanJose-spindephasing}, providing a strong 
motivation for  exploration.


In many cases, topological considerations 
entail robust phenomena, governed by some global properties, 
insensitive to microscopic details. 
The quantized Hall conductance arising in the quantum Hall effect
\cite{vonKlitzing,Thouless} is a  prime example. 
In this work, we address another robust  phenomenon\cite{Neumann,Herring}
rooted in topology, 
which appears in interacting spin systems subject to a magnetic field, 
the emergence of ground state degeneracies at certain magnetic fields.
In this case, a topological  invariant (an appropriately defined
global Chern number, to be referred here as the total
topological charge) predicts the existence and 
global properties of ground-state
\emph{magnetic degeneracy points}.\cite{Bruno,Gritsev,Berry,Garg,Roushan,Scherubl}
Most frequently, but not 
always, these degeneracy points are Weyl points\cite{Scherubl},
similar to linearly dispersing band touching points in the band structure of 
Weyl semimetals\cite{Armitage}, and also appearing  
in various physical contexts\cite{Riwar,Stenger,WenlongGao}.
The precise relation of the total topological charge, 
the magnetic degeneracy points,
and the topological charge of these points 
is explained in the context
 of molecular magnets  and spin-orbit-coupled double quantum dots
in Refs.~\onlinecite{Bruno} and \onlinecite{Scherubl}, respectively.

\begin{figure}
	\begin{center}
		\includegraphics[width=1\columnwidth]{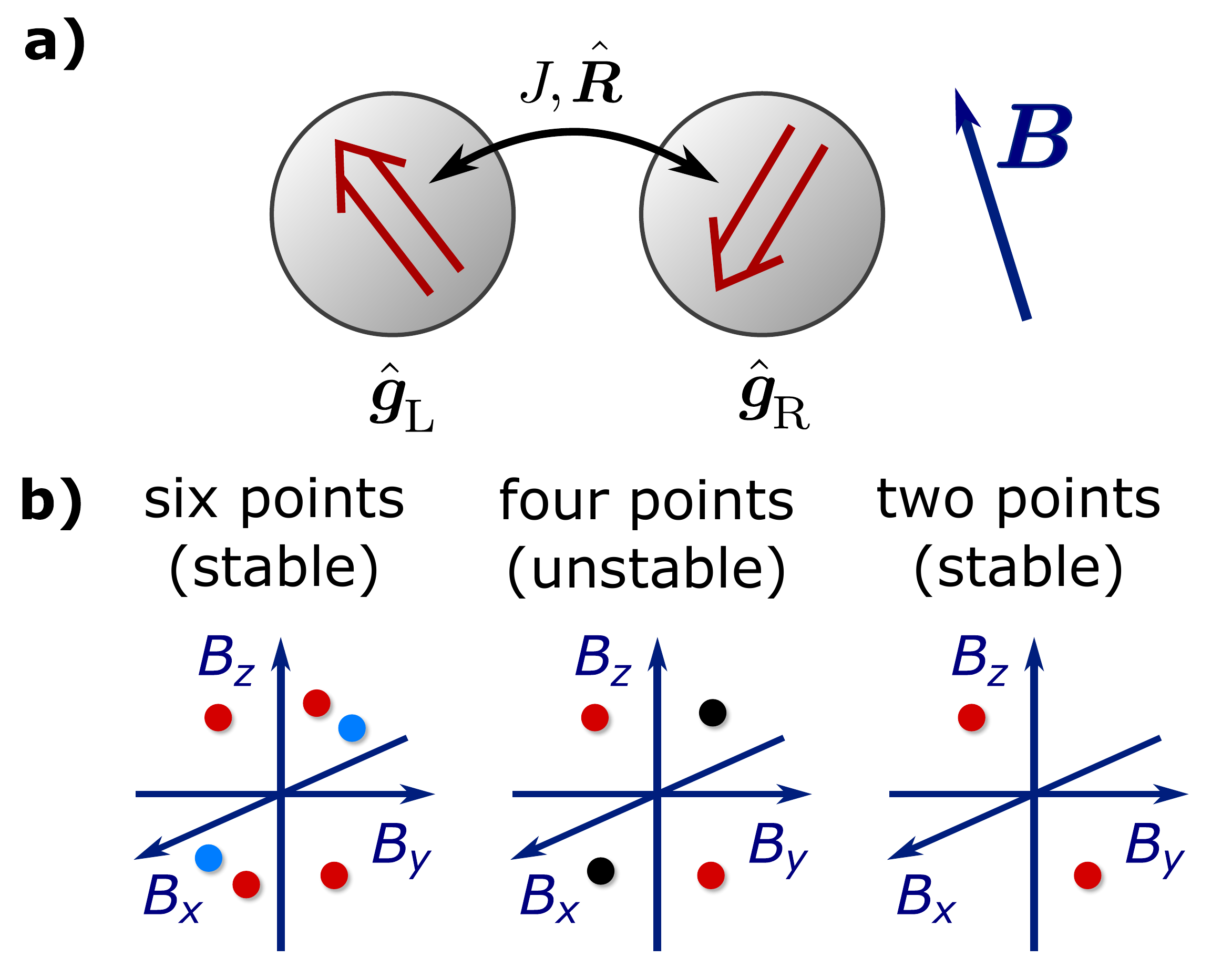}
	\end{center}
	\caption{\textbf{Magnetic degeneracy points of two interacting spin-1/2 electrons.}
	(a) Exchange interaction between the spins is described by its strength $J$ and a rotation $\hat{\vec R}$. External 
	magnetic field $\vec B$ couples to the spins through the $g$-tensors. 
	(b) Geometry and topological charge distribution of 
	magnetic degeneracy points for 
	$\text{det} \,\hat{\vec g}_\text{L},
	\text{det} \,\hat{\vec g}_\text{R} > 0$.
	Colors indicate topological charge:
	$+1$ (red), $-1$ (blue), $0$ (black).
	Six-point and  two-point patterns are stable,
	and the generic transition between these patterns is when 
	two oppositely charged pairs meet and their charges
	annihilate each other (four points).
	\label{fig:system}}
\end{figure}

Even though a nonzero total topological charge guarantees the
existence of magnetic degeneracy points, its value does not provide
a definite answer to the following questions. 
(i) What is the geometrical pattern
(isolated points, lines, surfaces, or their combinations)
drawn by the magnetic degeneracy points in the three-dimensional
magnetic parameter space?
(ii) How is the topological charge carried by the magnetic 
degeneracy points distributed among the points?
(iii) Are  different geometrical 
patterns and topological charge distributions stable against
small perturbations of the system's Hamiltonian?
These are nontrivial questions, and
answering them probably requires
extensive numerical investigations, in general. 


Here, we address questions (i), (ii), and (iii) for a specific, 
experimentally relevant setup, a spin-orbit-coupled 
interacting two-spin system\cite{Kavokin,Nowack-esr,NadjPerge-spinorbitqubit,NadjPerge,SchroerPRL2011,HarveyCollard,Tanttu,Scherubl}, 
and obtain exact results. 
We provide  a full classification of geometrical patterns
(and corresponding topological charge density patterns)
of the ground-state magnetic degeneracy points of this setup;
this `zoo' of patterns is introduced in Tables \ref{tab:table1}
and \ref{tab:table2}.
Finding the degeneracy points is reduced to the eigenvalue
problem of a $3 \times 3$ non-symmetric real matrix, and
hence our analysis inherits key features of different physics
subdisciplines, such as bifurcation theory\cite{Guckenheimer} and
non-Hermitian wave mechanics\cite{BerryNonhermitian,Heiss}, 
where similar eigenvalue problems play important roles. 
Beyond fundamental interest, the properties of 
magnetic degeneracy points are in fact practically important for 
control and readout of spin-based quantum bits; 
two-electron singlet-triplet degeneracy points, e.g.,  are often
exploited for spin initialization, control and 
readout\cite{Petta,Hanson,Reilly,Petta_beamsplitter,HarveyCollard,Tanttu}.

We consider a system of two interacting localized electrons,
subject to spin-orbit interaction,
and assume that they are placed 
in a homogeneous magnetic (Zeeman) field
(Fig.~\ref{fig:system}a). 
This system can be described by a $4\times 4$ Hamiltonian matrix\cite{Scherubl}
\bean
\label{eq:hamiltonian}
H &=& H_\text{Z} + H_\text{int}, \\
H_\text{Z} &=& \mu_\text{B} \vec B \left(\hat{
	\vec g}_\text{L} \vec S_\text{L} + \hat{\vec g}_\text{R} \vec S_\text{R}
\right), \\
H_\text{int} &=& J \vec S_\text{L} \hat{\vec R} \vec S_\text{R}.
\eean
Here, $H_\text{Z}$ is the Zeeman interaction with the external 
homogeneous magnetic field $\vec B$, 
$H_\text{int}$ is the spin-orbit-affected exchange interaction
between the two electrons, 
$\mu_\text{B}$ is the Bohr-magneton, 
$\hat{\vec g}_\text{L}$ and $\hat{\vec g}_\text{R}$ are the real-valued, 
spin-orbit-affected $g$-tensors
of the two electrons, $\vec S_\text{L}$ and $\vec S_\text{R}$ are the 
spin vector operators represented by $1/2$ times the spin-1/2 
Pauli matrices, $J>0$ is the strength of the exchange interaction, 
and $\hat{\vec R}$ is a  real, $3\times3$ special orthogonal matrix accounting for the spin-orbit
interaction in the exchange term.
The exchange term $H_\text{int}$ was 
derived from a two-site Hubbard model 
using quasidegenerate perturbation theory in 
Supplementary Note 2 of Ref.~\onlinecite{Scherubl},
see Eqs.~(S2) and (S3) therein, which imply 
that this term is a positive prefactor times 
a rotation.

The $g$-tensors are arbitrary real matrices, which are not
necessarily symmetric. 
For concreteness, let us first focus  on the case when 
the determinants of both $g$-tensors are positive,
$\text{det} (\hat{\vec g}_\text{L}), \,\text{det} ( \hat{\vec g}_\text{R}) > 0$. 
The elements of the three $3\times 3$ matrices 
$\hat{\vec g}_L$,
$\hat{\vec g}_R$, and $\hat{\vec{R}}$
are determined
by microscopic details (spin-orbit interaction, confinement potential, etc.),
but here we treat them as possibly independent phenomenological
parameters. 

In our topological considerations, we distinguish the three Cartesian magnetic-field components $B_x$, $B_y$, $B_z$ in the 
Hamiltonian as `primary' parameters, and refer to further parameters as
`secondary' ones.
In this nomenclature, secondary parameters are fixed, while primary parameters are thought of as external parameters, varied continuously. At certain points within the space of primary parameters, the ground state of $H$ becomes degenerate. 
We refer to these points as \emph{magnetic degeneracy points}.

We have studied this two-spin system  in detail in our recent work~\cite{Scherubl}.
There we have shown that in the case
$\text{det} (\hat{\vec g}_\text{L}), \,\text{det} ( \hat{\vec g}_\text{R}) > 0$,
(i) topological considerations guarantee the existence of  ground-state magnetic degeneracy points
(often, but not always, magnetic Weyl points),  irrespective of microscopic  details of the Hamiltonian, 
(ii) the degeneracy points  carry topological charge, and the total  topological charge carried 
by all degeneracy points of the 
three-dimensional magnetic-field parameter space 
sums up to +2.  By numerical work and intuitive considerations, 
we have demonstrated four different  geometrical patterns formed by the magnetic degeneracy points, 
and the corresponding topological charge distributions: (A) A sphere, carrying a
surface topological charge of +2. This is the case without spin-orbit interaction, when the
$g$-tensors are isotropic and the exchange interaction is
of antiferromagnetic Heisenberg type ($J\vec{S}_{\text{L}}\cdot\vec{S}_{\text{R}}$).  This case also provides an example where the magnetic degeneracy points
are not Weyl points.  (B) Two isolated (Weyl) points, each carrying charge +1.
(C) Six isolated (Weyl) points, four of them carrying charge +1, two of them carrying 
charge -1. (D) Four isolated points, two of them carrying charge +1, the other two 
carrying no charge. (See Fig.~\ref{fig:system}b  of this work or Fig.~4a of Ref.~\onlinecite{Scherubl}.)

We note that we use the term Weyl point to label an
isolated degeneracy point that possesses all of the following properties:
(1)  the energy splitting in its vicinity is linearly increases with small deviation in the parameter space, in all directions, 
(2) the degeneracy is twofold, 
and
(3) the absolute value of its topological charge is one. 
Accordingly, we did not label the charge-neutral degeneracy points
in case (D) above as Weyl-points, since they violate both (1) and 
(3). 
In the condensed-matter literature, band degeneracy points 
not showing at least one of the above three properties are sometimes
called multi-Weyl  
points\cite{ChenFang_multiweyl,ZhongboYan,Ahn,ZeMinHuang}; we do not 
use this terminology here. 

Going beyond our earlier numerical study,
here we develop an analytical approach, which allows for a complete 
classification of  the geometrical and topological 
structure of magnetic degeneracy points. 
For this analysis, it is important to distinguish 
three cases defined by the three possible values of 
the total topological charge  $\cal Q$. 
This charge
is the sum of the $g$-tensor determinants, i.e., 
${\cal Q}=
\text{sgn(det}\ \hat{\vec g}_{\text{L}})+\text{sgn(det}\ \hat{\vec g}_{\text{R}})$,
and hence ${\cal Q} \in \{-2, 0, 2\}$
We will show that the geometrical patterns are the same
in case ${\cal Q} = +2$ (Table \ref{tab:table1}) 
and case ${\cal Q} = -2$, although the 
topological charges are the opposite in the two cases.
Also, we will show that the geometrical patterns are distinct 
in the case ${\cal Q} = 0$ (Table \ref{tab:table2}).
Rephrasing this in terms of the relative sign
${\cal S}=
\text{sgn}[\text{det} (\hat{\vec g}_\text{L})\cdot
\text{det}(\hat{\vec g}_\text{R})]$,
we can say that the geometrical patterns are different for 
the two possible values ${\cal S} \in \{-1,+1\}$.

For ${\cal S} = +1 $ we obtain 
a  sixfold classification of geometrical patterns of
magnetic degeneracy points (Table \ref{tab:table1}), corresponding to 
six different topological  
charge distributions (i.e., two more beyond the ones identified in Ref.~\onlinecite{Scherubl}), 
and further four possible classes are identified for  ${\cal S} = -1$.
These form altogether  \emph{ten geometrical classes}. 

Furthermore, we use our approach to characterize the 
stability of these patterns against small perturbations of the Hamiltonian
(see Fig.~\ref{fig:system}b): 
we find two stable  and one almost stable charge configurations both for 
${\cal S} = +1$ and 
for   ${\cal S} = -1$,
 the almost stable magnetic configuration forming a generic boundary between stable configurations,
similar to bifurcations in the theory of dynamical systems.\cite{Guckenheimer}

The rest of the paper is structured as follows.
In Sec.~\ref{sec:classification}, we derive a classification of the
geometrical patterns and topological charge density patterns
of magnetic degeneracy points for the case ${\cal S} =+1$, 
with the main results summarized in Table \ref{tab:table1}.
In Sec.~\ref{sec:stability}, we analyze the stability of each of these
patterns against perturbations of the Hamiltonian.
In Sec.~\ref{sec:det_min}, we extend the results to
case ${\cal S} = -1$, and provide our conclusions 
in Sec.~\ref{sec:conclusions}.

\section{Classifications of degeneracy points}
\label{sec:classification}

\subsection{Mapping the degeneracy problem to the eigenproblem
of a non-symmetric matrix}
Given the Hamiltonian \eqref{eq:hamiltonian}, 
it is not obvious how to analyze the geometrical patterns
formed by the magnetic degeneracy points. 
One could, in principle, 
find the eigenvalues of the  $4\times4$ Hamiltonian \eqref{eq:hamiltonian}
analytically, 
and investigate, in a very large dimensional parameter space, the conditions 
under which ground-state degeneracies occur. 
Numerical diagonalization and numerical search for 
the magnetic degeneracy points is also an option, but this requires a lot of 
computational effort and is most likely incomplete~\cite{Scherubl}. 

Fortunately, for the specific  Hamiltonian, Eq.~\eqref{eq:hamiltonian},
a simple observation allows for a fully 
analytical treatment of the problem.  As a first step, we introduce a local spin transformation that
leaves the left spin invariant, $\vec S'_\text{L} = \vec S_\text{L}$, but
it rotates the right spin with the rotation appearing in the 
exchange term, 
$\vec S'_R = \hat{\vec R} \vec S_R$.
This transformation renders the exchange interaction isotropic. 
The transformed Hamiltonian now reads 
\bean\label{eq:transformed}
H' = \mu_\text{B} \left(
\vec B_\text{L,eff} \vec S'_\text{L}  +
\vec B_\text{R,eff} \vec S'_\text{R}
\right) + J \vec S'_\text{L} \vec S'_\text{R},
\eean
where the effective magnetic fields felt by the left and right spins read 
$\vec B_\text{L,eff} = \vec B \hat{\vec g}_\text{L}$ 
and 
$\vec B_\text{R,eff} = \vec B \hat{\vec g}_\text{R} \hat{\vec R}^{-1}$. 
Clearly, if these effective magnetic fields are parallel, then we can take the 
spin quantization axis $z$ along the direction of the effective magnetic fields.
This choice implies that $S'_{\text{L}z}+S'_{\text{R}z}$ is
conserved, and therefore,  if the effective fields point in the same direction, then
there must be a ground-state level crossing as $B$ is increased from zero to infinity along this direction (Appendix~\ref{app:locations}).

These observations imply that for finding the magnetic
degeneracy points, we do not have to solve the
eigenvalue problem of the $4 \times 4$ Hamiltonian, 
but it is instead sufficient to find the magnetic-field directions
for which $\vec B_\text{L,eff} \parallel \vec B_\text{R,eff}$. 
An elementary proof shows that this condition is satisfied, 
if and only if $\vec B$ is a (left) eigenvector
of the $3 \times 3$ matrix 
\bean
\label{eq:matrixm}
\hat{\vec M} = \hat{\vec g}_\text{L} \hat{\vec R} \hat{\vec g}_\text{R}^{-1}.
\eean
A  magnetic degeneracy point is found in this direction if
 the eigenvector belongs to a positive  eigenvalue of $\hat{\vec M}$.
Note that $\hat{\vec M}$ is real, but in general it is not symmetric.

\begin{table*} [htbp]
	\centering
	\begin{tabular}{||c|c|c|c|c|c|c||}
		\hline \thead{Eigenpattern\\ label}& \thead{Multiplicities of\\eigenvalues}& \thead{Geometric\\ pattern}&\thead{Topological charge\\ distribution}&\thead{Degeneracy\\ points}&\thead{Jordan\\ normal form}&\thead{Stability\\ codimension}\\
		\hline \hline
		(I)&(1/1,1/1,1/1)&six points&$4\times(+1)$, $2\times (-1)$&
		\begin{minipage}{.15\textwidth}
			\includegraphics[width=0.9\linewidth]{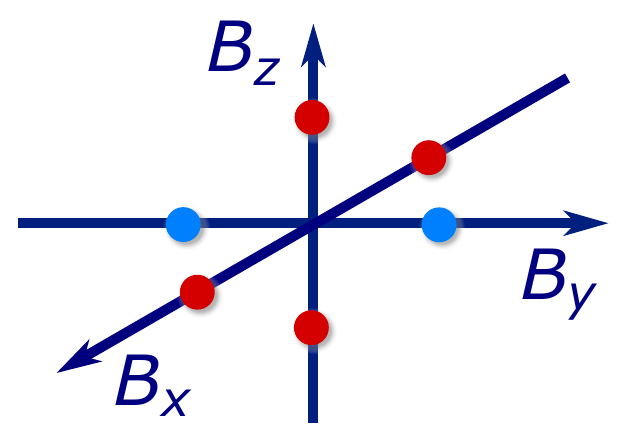}
		\end{minipage}
		&
		$\begin{pmatrix}
		a&0&0\\
		0&b&0\\
		0&0&c\\ 
		\end{pmatrix}$&0 (stable)\\
		\hline
		(II)&(2/2,1/1)&two points, loop&$2\times(+1)$, $0$&
		\begin{minipage}{.15\textwidth}
			\includegraphics[width=0.9\linewidth]{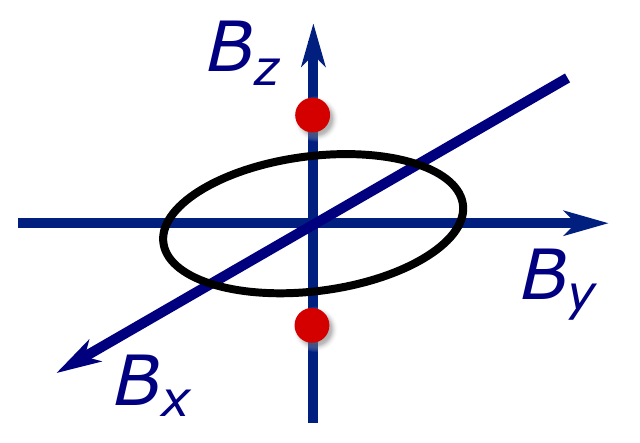}
		\end{minipage}
		&
		$\begin{pmatrix}
		a&0&0\\
		0&a&0\\
		0&0&b\\ 
		\end{pmatrix}$&3\\
		\hline
		(III)&(2/1,1/1)&four points&$2\times(+1)$, $2\times 0$&
		\begin{minipage}{.15\textwidth}
			\includegraphics[width=0.9\linewidth]{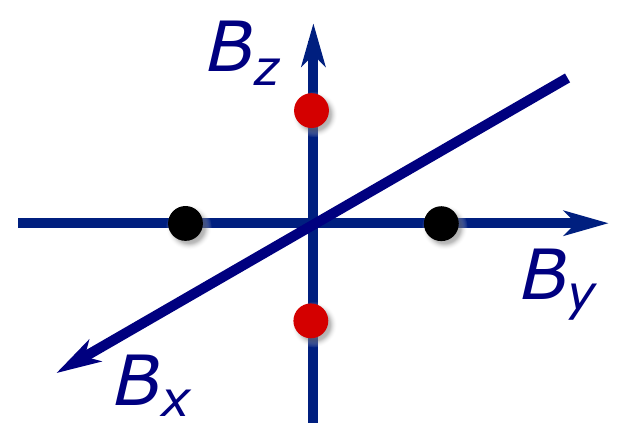}
		\end{minipage}
		&
		$\begin{pmatrix}
		a&1&0\\
		0&a&0\\
		0&0&b\\ 
		\end{pmatrix}$&1 (almost stable)\\
		\hline
		(IV)&(3/3)&closed surface&$2$ (surface charge)&
		\begin{minipage}{.15\textwidth}
			\includegraphics[width=0.9\linewidth]{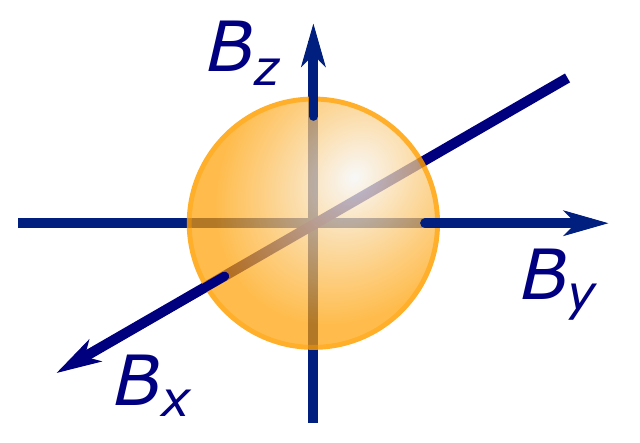}
		\end{minipage}
		&
		$\begin{pmatrix}
		a&0&0\\
		0&a&0\\
		0&0&a\\ 
		\end{pmatrix}$&8\\
		\hline
		(V)&(3/2)&loop&$2$ (line charge)&
		\begin{minipage}{.15\textwidth}
			\includegraphics[width=0.9\linewidth]{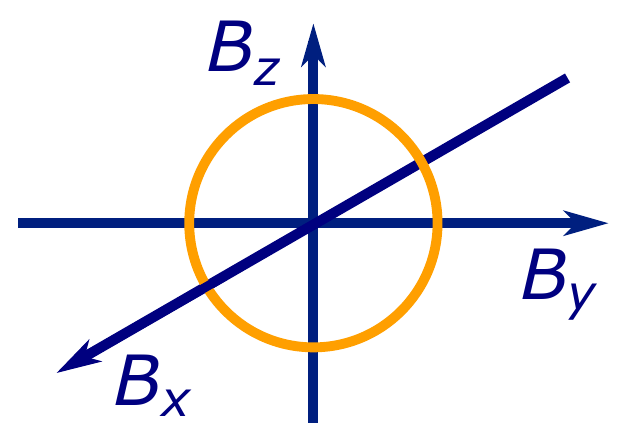}
		\end{minipage}
		&
		$\begin{pmatrix}
		a&1&0\\
		0&a&0\\
		0&0&a\\ 
		\end{pmatrix}$&4\\
		\hline
		(VI)&(3/1)&two points&$2\times(+1)$&
		\begin{minipage}{.15\textwidth}
			\includegraphics[width=0.9\linewidth]{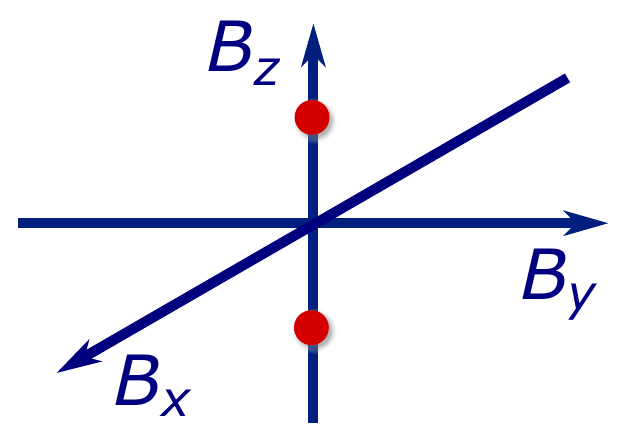}
		\end{minipage}
		&
		$\begin{pmatrix}
		a&1&0\\
		0&a&1\\
		0&0&a\\ 
		\end{pmatrix}$&2\\
		\hline
		(VII)&(1/1)&two points&$2\times(+1)$&
		\begin{minipage}{.15\textwidth}
			\includegraphics[width=0.9\linewidth]{zoo1.pdf}
		\end{minipage}
		&
		$\begin{pmatrix}
		a&0&0\\
		0&\text{n/c}& 0/1\\
		0&0&\text{n/c}\\ 
		\end{pmatrix}$&0 (stable)\\
		\hline
	\end{tabular}
	\caption{
	{\bf Zoo of geometrical patterns and topological charge density 
	patterns of magnetic ground-state degeneracies of a spin-orbit-coupled
	two-spin system.}
	Classification is based on the eigenpatterns of the $3 \times 3$ real matrix $\hat{\vec M}$ with positive determinant, see text. In the second column, we display the algebraic and geometrical multiplicities of positive eigenvalues.
	Each eigenpattern implies a geometrical pattern of the magnetic degeneracy points, which in turn implies a topological charge density pattern. 
The fifth column schematically shows the geometrical pattern
of degeneracy points, 
and the topological charge of each geometrical element:
$+2$ (orange), $+1$ (red), $0$ (magenta), $-1$ (blue).
In sixth row, the Jordan normal form may have
negative or complex eigenvalues (`n/c'), 
and if there is only a single negative eigenvalue, then 
the superdiagonal element of its Jordan block can be
either 0 or 1. The last column shows the stability codimension of the eigenpatterns, i.e., the number of linearly independent constraints for a small perturbation not to break the pattern. 
\label{tab:table1}}
\end{table*}

\subsection{Eigenpattern of the matrix $\hat{\vec M}$ and  the geometrical pattern of the magnetic degeneracy points}

As we now show, the 
structure of the eigenvalues and eigenvectors 
of the matrix $\hat{\vec M}$ implies 
the geometrical pattern of the magnetic degeneracy points.
For concreteness, throughout the rest of Sec.~\ref{sec:classification} and also 
in Sec.~\ref{sec:stability}, we focus on the case
where the determinants of both
$g$-tensors are positive, 
$\text{det} (\hat{\vec g}_\text{L}), \,\text{det} ( \hat{\vec g}_\text{R}) > 0$,
that is, ${\cal S} = +1$ and ${\cal Q}=+2$. 
The results obtained trivially carry over to the case 
$\det(\hat{\vec g}_\text{L}), \det(\hat{\vec g}_\text{R}) < 0$ (${\cal S} = +1$, ${\cal Q}=-2$), 
with the only modification that the total topological charge of the ground-state
magnetic degeneracy points has opposite sign in the latter case, 
and correspondingly, all topological charges  in the discussions below are reversed.

The case ${\cal S} = -1$ (${\cal Q}=0$) is, however, 
quite different. Then the total topological charge of the magnetic degeneracy 
points adds up to zero, 
implying a completely different structure of the eigenproblem of
$\hat{\vec M}$, and thereby different geometrical patterns and
topological charge distributions.
This  case will be discussed in Sec.~\ref{sec:det_min}.


For $\text{det} (\hat{\vec g}_\text{L}), \,\text{det} ( \hat{\vec g}_\text{R}) > 0$,
 the $3 \times 3$ matrix $\hat{\vec M}$ is real but non-symmetric in general,
and it has a positive determinant, because the $g$-tensors have positive
determinants, and $\text{det} \, \hat{\vec R} =1$.
To see the possible geometries of the magnetic degeneracy points, we first 
have to identify qualitatively different solutions of the eigenproblem of 
$\hat{\vec M}$. 

As we show below, 
for $\text{det} (\hat{\vec g}_\text{L}), \,\text{det} ( \hat{\vec g}_\text{R}) > 0$
there are 7 different 
cases, labelled from (I) to (VII) in Table~\ref{tab:table1},
that are 
classified by the number of positive
eigenvalues, and their algebraic and geometric multiplicities.
(For a brief summary of the relevant concepts and relations of 
linear algebra, see Appendix \ref{app:linalg}.)
We call these cases the \emph{eigenpatterns} of the
matrix $\hat{\vec M}$, and will use, e.g., 
$\text{ep}(\hat{\vec M}) = \text{(IV)}$, to 
denote that the eigenpattern of $\hat{\vec M}$ is (IV).

For $\text{ep}(\hat{\vec M}) = \text{(I)}$, the matrix
$\hat{\vec M}$ has three different positive eigenvalues,
$a$, $b$ and $c$.
The algebraic and geometric multiplicities are all 1, 
as denoted in the second 
column
of Table \ref{tab:table1}.
Let us denote the normalized left 
eigenvectors with $\vec v_a$, $\vec v_b$ and $\vec v_c$. 
Then, there are six magnetic degeneracy points, 
one time-reversed pair associated to each eigenvector $\vec v_\alpha$, 
appearing if the magnetic 
field is aligned or antialigned with those eigenvectors. 
The locations of the degeneracy points in the
original magnetic-field parameter space are
\bean
\vec{B}_{\alpha,\pm} = \pm B_{\alpha} \cdot \vec v_{\alpha}, \, \, \,
(\alpha \in a,b,c).
\eean
The expressions for the critical fields $B_{\alpha}$,
and their derivations are summarized in
Appendix~\ref{app:locations}.
Topological charges of these degeneracy points are discussed below.

Different eigenpatterns arise when $\hat{\vec M}$ has 
two different positive eigenvalues, $a$ and $b$,
see Table \ref{tab:table1}, rows (II) and (III).
In these cases, one of these eigenvalues, say $a$, 
must have an algebraic multiplicity of 2. 
(Otherwise, either $\hat{\vec M}$ would have a third, negative eigenvalue, 
which is 
forbidden by the fact the $\hat{\vec M}$ has positive determinant, 
or it would have a 
third, complex eigenvalue, which is forbidden by the fact that complex 
eigenvalues come in complex-conjugate pairs.) 
Then, $a$ can have a geometric multiplicity of 2, yielding case (II)
in Table~\ref{tab:table1},
or a geometric multiplicity of 1, yielding case (III). 
In case (II), the magnetic 
degeneracy points are arranged at two isolated points along the line of 
$\vec v_{b}$, at $\vec B_{b,\pm} = \pm B_{b}\cdot\vec {v}_{b}$, 
and along an ellipse in the plane spanned by the
two remaning eigenvectors, 
$\vec v_{a_1}$ and $\vec v_{a_2}$.
See Appendix~\ref{app:ellipse} for details.
In case (III), the magnetic degeneracy points are arranged in four isolated points, similarly at $\vec B_{a,\pm} = \pm B_{a}\cdot\vec {v}_{a}$ and 
$\vec B_{b,\pm} = \pm B_{b}\cdot\vec {v}_{b}$.

Further eigenpatterns appear when $\hat{\vec M}$ has a single positive 
eigenvalue, $a > 0$,
see rows (IV)-(VII) of Table \ref{tab:table1}. 
These eigenvalues 
can have an algebraic multiplicity of 3, and a geometric multiplicity of 3,
yielding case (IV), where the magnetic degeneracy points form a closed 
surface, an ellipsoid. 
The simplest example is the case without spin-orbit interaction, 
where the $g$-tensors and $\hat{\vec R}$ are all proportional
to the unit matrix, and the magnetic degeneracy points form a
sphere.
Alternatively, $a$ can have an algebraic multiplicity of 3, and a 
geometric multiplicity of 2, yielding case (V), where the magnetic 
degeneracy points form a closed loop, an ellipse. 
Yet another alternative is that $a$ has a geometric multiplicity of 1, 
yielding case (VI), with two isolated magnetic degeneracy points. 
Finally, it can also have an algebraic multiplicity of 1, and consequently a 
geometric multiplicity of 1, denoted as case (VII), yielding two isolated
 magnetic degeneracy points.
This sevenfold classification of the relevant solutions of the eigenvalue 
problem implies a sixfold classification of the qualitatively different 
geometrical  patterns of the magnetic degeneracy points, since 
the eigenpatterns (VI) and (VII) yield  the same geometry.

\subsection{Topological charge-density patterns}
We know that the magnetic degeneracy points can also carry a topological 
charge \cite{Scherubl},
and for our Hamiltonian with fixed secondary parameters, the total 
charge is ${\cal Q} = +2$. 
In principle, it could happen that two different Hamiltonians yield the same 
geometry of degeneracy points, but the charge is distributed differently 
among the elements. 
For example, in case (II), the ellipse-shaped loop 
could be neutral and the isolated points 
could carry charge +1 each, or the points could be neutral and the ellipse 
could carry charge +2. 
As we show in Appendix~\ref{app:charge}, the topological
charge density is uniquely determined by the geometrical
pattern;
in the example above, the points are charged and the loop
is neutral. 
The topological charge density patterns are listed in the fourth
column and sketched in the fifth column of Table~\ref{tab:table1}.

\section{Stability analysis of eigenpatterns and corresponding  geometrical patterns}
\label{sec:stability}

In Ref.~\onlinecite{Scherubl}, we have studied random Hamiltonians for this 
spin-orbit-coupled two-spin model numerically, and among those random Hamiltonians, 
we have identified only two of the above six different geometrical patterns. 
Why don't we find representatives of the other four geometrical patterns
in a random ensemble of Hamiltonians?
As we argue below, each eigenpattern 
can be 
characterized by a `degree of stability' or `codimension', 
denoted by $d$, 
which is a non-negative integer, familiar from the codimension
property of bifurcations\cite{Guckenheimer}: 
if $d=0$, then the eigenpattern is stable, 
if $d>0$, then the eigenpattern is unstable, 
and an increasing $d$ is interpreted as decreasing stability.
 
We define stability via sensitivity to small random perturbations. 
Consider the 
Hamiltonian $H$ of Eq.~\eqref{eq:hamiltonian} 
with fixed secondary parameters, 
which specifies the matrix $\hat{\vec M}$, which in turn  
has a specific eigenpattern.
If we slightly modify the secondary 
parameters, and thereby add 
an infinitesimal perturbation, $ H\to H' = H + \delta H$,  then
the eigenpattern of $H'$ may be the same as that of $H$, 
or it may be different. 
If the eigenpattern of $H'$ is the same as that of $H$ for any
infinitesimal perturbation $\delta H$, then we call the eigenpattern
of $H$ stable. 
Otherwise, we call it unstable. 

Instead of considering $H$ directly, we address
the question of eigenpattern stability 
by regarding the matrix $\hat{\vec M}$ as the element of 
a 9-dimensional vector space. 
\footnote{We can modify all elements of $\hat{\vec M}$ by changing some parameters of 
the original Hamiltonian.}
The infinitesimal perturbations $\delta  \hat{\vec M}$ span
a 9-dimensional vector space, too; we denote this vector space
by $W$. 
The question of eigenpattern stability can then be phrased as follows: 
for a given $\hat{\vec M}$, 
what is the dimension of the subspace $W_\text{s} \leq W$
spanned by the infinitesimal perturbations $\delta \hat{\vec M}$ 
that preserve the eigenpattern of $\vec{\hat M}$ under
$\vec{\hat M} \to \vec{\hat M}' = \vec{\hat M} + \delta \vec{\hat M}$ ?
If $\text{dim} \left(W_\text{s} \right) = 9$, then 
the eigenpattern is preserved for an arbitrary infinitesimal perturbation, 
i.e., the eigenpattern of $\vec{\hat M}$ is stable. 
Otherwise, it is unstable, and the degree of stability 
can be characterized by the codimension of the stable
subspace $W_\text{s}$, which is 
$d \equiv 9- \text{dim} \left(W_\text{s}\right)$, 
with $d=0$ denoting the stable case and increasing 
$d$ signalling increasing instability.

We now outline a method  to calculate $d$ for a given $\hat{\vec M}$. 
This is based on the Jordan decomposition  of $\hat{\vec M}$  (see Appendix \ref{app:linalg}),
 \bean
 \label{eq:jordandecomposition}
 \hat{\vec M} = \hat{\vec P} \hat{\vec J} \hat{\vec P}^{-1},
 \eean
 with
 $\hat{\vec J}$ being the Jordan normal form 
 of $\hat{\vec M}$ and
 $\hat{\vec P}$ 
 a similarity transformation.
 Let us choose  $\text{ep}(\hat{\vec M}) = $ (V) as our example, 
so its Jordan normal form reads
\bean
\hat{\vec J} = \left(\bna{ccc}
\lambda & 1 & 0 \\
0 & \lambda & 0 \\
0 & 0 & \lambda
\eda \right).
\eean
The matrix $\hat{\vec M}$ has thus one eigenvalue with an algebraic multiplicity of 3 but 
only two linearly independent corresponding eigenvectors. 
Recall that this eigenpattern implies that the
magnetic degeneracy points are located on an ellipse.

We first characterize those perturbations of $\hat{\vec M}$ which 
 preserve this eigenpattern, that is, preserve  the structure of the Jordan form. For these,  the Jordan form of the deformed matrix $\hat{\vec M}'$ must
read
\bean
\hat{\vec J}' = \hat{\vec J} + \delta \hat{\vec J}
=
\hat{\vec J} + \left(\bna{ccc}
\nu & 0 & 0 \\
0 & \nu & 0 \\
0 & 0 & \nu
\eda \right),
\eean
with an infinitesimal $\nu$.
Since the only constraint on the perturbation is that
the eigenpattern (that is, the Jordan normal form) should be 
preserved, an arbitrary infinitesimal change is allowed in the
similarity transformation $\hat{\vec P}$,
\bean
\hat{\vec P}' = \hat{\vec P}(\mathds{1} + \delta\hat{\vec B})
\eean
with an infinitesimal term
\bean
\delta \hat{\vec B} = 
\left(\bna{ccc}
b_{11} & b_{12} & b_{13} \\
b_{21} & b_{22} & b_{23} \\
b_{31} & b_{32} & b_{33} 
\eda \right).
\eean

Now we have parametrized, using 10 infinitesimal parameters
(the $b_{ij}$-s and $\nu$), 
all matrices that are infinitesimally 
close to $\hat{\vec M}$ and  have the same
eigenpattern as $\hat{\vec M}$;
in fact, we have overparametrized $\hat{\vec M}$.
We can express the shift of the matrix
$\hat{\vec M}$ 
up to linear order in these infinitesimal parameters as
\bean
\delta \hat{\vec M} &=& \hat{\vec M}' - \hat{\vec M} = 
\hat{\vec P}' \hat{\vec J}' \hat{\vec P}'^{-1} - 
\hat{\vec P} \hat{\vec J} \hat{\vec P}^{-1} \nonumber \\
&=& 
\hat{\vec P}\left(
[\delta \hat{\vec B},\hat{\vec J}] + \delta \hat{\vec J}
\right) \hat{\vec P}^{-1}
\equiv
\hat{\vec P} \, \delta \! \hat{\widetilde{\vec M}} \, \hat{\vec P}^{-1}.
\label{eq:sim}
\eean

Not all our infinitesimal parameters lead, however, to independent deformations of $\hat M$.
To determine  \emph{independent} deformations, we note
that $\delta \! \hat{\widetilde{\vec M}} = \delta \! \hat{\widetilde{\vec M}}(\{b_{ij}\},\nu)$ 
is a homogenous linear 
function of the infinitesimal parameters, that is,
\bean
\label{eq:linear}
\delta \! \hat{\widetilde{\vec M}}_{ij} = \sum_{\gamma = 1}^{10} C_{ij,\gamma} \epsilon_\gamma,
\, \, \, (i,j = 1, 2, 3),
\eean
where $\left(\{\epsilon_\gamma\}\right) = \left(\{b_{ij}\},\nu\right)$ 
is the 10-tuple
of the infinitesimal parameters,
and the coefficients of the linear relation 
are arranged in  the $9 \times 10$ matrix $C$.
This linear relation Eq.~\eqref{eq:linear} together with the similarity transformation \eqref{eq:sim}
implies that the dimension of the image $\delta \hat {\vec M}$ of the 
10-dimensional vector space of the infinitesimal parameters
is simply $\text{rank}(C)$. 
The dimension of the stable subspace of perturbations
is therefore $\text{dim}\left( W_\text{s} \right) =\text{rank}(C)$.

A straightforward calculation shows that in this specific case, 
$\text{dim}\left( W_\text{s} \right)=\text{rank}(C) = 5$.
Correspondingly, the stability codimension is
$d = 9 - \text{dim}\left( W_\text{s} \right) = 4$ 
for eigenpattern (V), cf.~Table \ref{tab:table1}. 
We therefore conclude that the eigenpattern (V) has a 
rather high codimension $d$, and is therefore quite unstable.

The stability of each of the 7 eigenpatterns in Table~\ref{tab:table1}
can be characterized by calculating the corresponding
codimension $d$ in a similar way.
The results are shown in the seventh column of Table~\ref{tab:table1}.
The most important result is that the stability codimension of
eigenpatterns (I) and (VII) are zero, hence these are the 
stable eigenpatterns, and consequently, these provide
two geometrical patterns of the magnetic 
degeneracy points: the `two points' configuration (VII), 
and the `six points' configuration (I). 
This result explains and corroborates our earlier numerical 
finding \cite{Scherubl}, where only these two geometrical
patterns were found by studying randomized Hamiltonians.

Note that the `six points' geometrical pattern is always stable;
however, the `two points' configuration can also be
unstable, when the eigenpattern (VI) is realized. 
In fact, when the Hamiltonian belonging to eigenpattern (VI) 
is subject to an arbitrary infinitesimal 
perturbation $\delta \hat{\vec M}$, then it might 
(i) preserve its eigenpattern (if $\delta \hat{\vec M} \in W_\text{s}$), or
(ii) change its eigenpattern to (VII), or
(iii) change its eigenpattern to (I), i.e., the two degeneracy points 
can split into three pairs of Weyl points, or
(iv) change its eigenpattern to (III), i.e., the two degeneracy points split into a Weyl and a neutral point pair.

It is also remarkable that the textbook case of a spherical surface geometry 
of the magnetic degeneracy points, provided by isotropic $g$-factors
and isotropic Heisenberg interaction, is the \emph{most unstable} 
of all configurations: its stability codimension $d = 8$ is maximal 
among the seven eigenpatterns.

A further question is, 
how transitions between stable eigenpatterns 
take place upon  changing the secondary parameters of the Hamiltonian?
If we consider two Hamiltonians from the two different
stable eigenpattern classes (I) and (VII), and continuously 
interpolate between them, then there must be a critical 
point on the way where four of the six points disappear. 
The answer is given by Table~\ref{tab:table1}, and
also depicted in Fig.~\ref{fig:system}b: the only
geometric pattern with a stability codimension 1 is 
the `four points' pattern, hence this is the generic
boundary between the two stable geometric patterns.
To reach the remaining four patterns, further fine tuning is 
required.

\begin{table*} [htbp]
	\centering
	\begin{tabular}{||c|c|c|c|c|c|c||}
		\hline \thead{Eigenpattern\\ label}& \thead{Description}& \thead{Geometric\\ pattern}&\thead{Topological charge\\ distribution}&\thead{Degeneracy\\ points}&\thead{Jordan\\ normal form}&\thead{Stability\\ codimension}\\
		\hline \hline
		(VIII)&(1/1,1/1)&four points&$2\times(+1)$, $2\times (-1)$&
		\begin{minipage}{.15\textwidth}
			\includegraphics[width=0.9\linewidth]{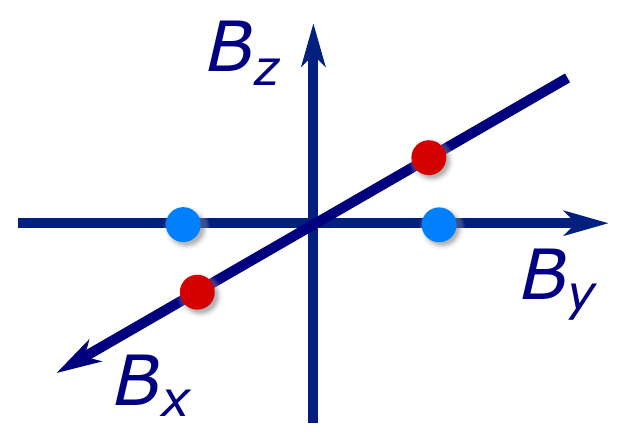}
		\end{minipage}
		&
		$\begin{pmatrix}
		a&0&0\\
		0&b&0\\
		0&0&\text{neg}\\ 
		\end{pmatrix}$&0 (stable)\\
		\hline
		(IX)&(2/2)&loop&$0$&
		\begin{minipage}{.15\textwidth}
			\includegraphics[width=0.9\linewidth]{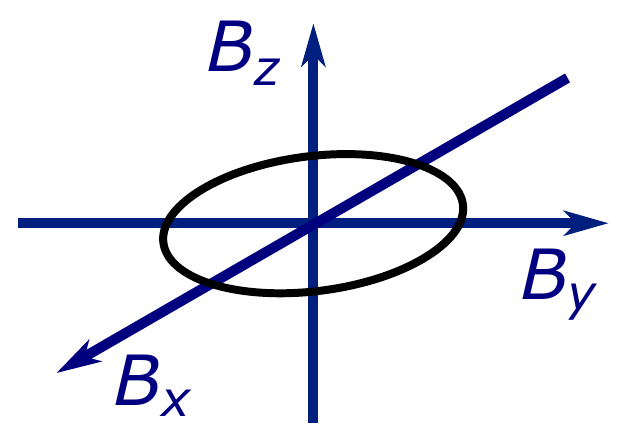}
		\end{minipage}
		&
		$\begin{pmatrix}
		a&0&0\\
		0&a&0\\
		0&0&\text{neg}\\ 
		\end{pmatrix}$&3\\
		\hline
		(X)&(2/1)&two points&$2\times 0$&
		\begin{minipage}{.15\textwidth}
			\includegraphics[width=0.9\linewidth]{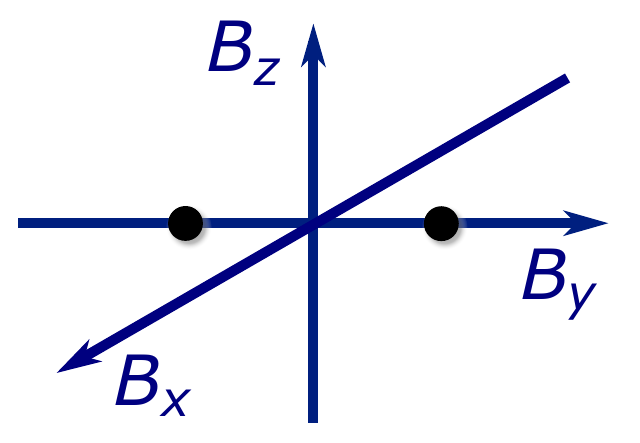}
		\end{minipage}
		&
		$\begin{pmatrix}
		a&1&0\\
		0&a&0\\
		0&0&\text{neg}\\ 
		\end{pmatrix}$&1 (almost stable)\\
		\hline
		(XI)&(-)&no points&no charge&
		\begin{minipage}{.15\textwidth}
			\includegraphics[width=0.9\linewidth]{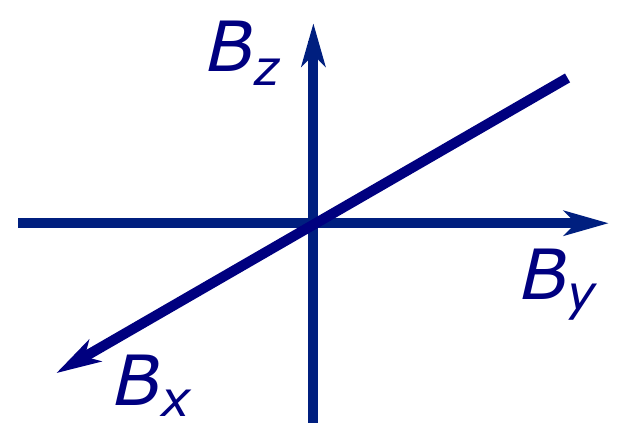}
		\end{minipage}
		&
		$\begin{pmatrix}
		\text{n/c}&0/1&0\\
		0&\text{n/c}&0/1\\
		0&0&\text{neg}\\ 
		\end{pmatrix}$&0 (stable)\\
		\hline
	\end{tabular}
	\caption{
	{\bf Zoo of geometrical patterns and topological charge density 
	patterns if the $g$-tensors have opposite signs.}
	Notation and the structure of the table 
	follows that in Table \ref{tab:table1}. 
	The label `neg' refers to a negative eigenvalue.
In eleventh row, the Jordan normal form may have
negative or complex eigenvalues (`n/c').
If two or three negative eigenvalues coincide, 
then the superdiagonal elements of the 
corresponding Jordan block can be
either 0 or 1.
 \label{tab:table2}}
\end{table*}

\section{Patterns of magnetic degeneracy points for ${\cal S} = -1$}
\label{sec:det_min}

Patterns of magnetic degeneracy points appearing for
a negative relative sign ${\cal S} = -1$ of the $g$-tensors
are different from the ones discussed in the previous two 
sections.
In this case, the total topological charge of the ground state magnetic degeneracy points is ${\cal Q}=0$. 
This indicates that generic Hamiltonians could exist 
without \emph{any}  magnetic degeneracy. Also, since Weyl points
must appear in $\pm \vec B$ pairs, and their charge must add up to 0, 
we  expect another generic situations with four magnetic Weyl points, two with topological charge $+1$ and two with 
topological charge $-1$. 

These expectations 
are indeed confirmed by the eigenpattern analysis of the matrix $\hat{ \vec M}$.
The analysis follows the steps of the previous section. 
The matrix $\hat{ \vec M}$ defined in Eq.~\eqref{eq:matrixm}
has a negative determinant in this case, but still 
the magnetic degeneracy points are associated with the positive
eigenvalues of $\hat{\vec M}$. The negative determinant of $\hat{\vec M}$ implies that 
the combinations of algebraic and geometric
multiplicities of its eigenvalues are different from the 
positive-determinant case in the main text. 
Apart from these differences, the analysis is very similar, 
hence we omit the details here, and 
summarize the results in Table~\ref{tab:table2}.

\begin{figure}[h]
	\begin{center}
		\includegraphics[width=0.9\columnwidth]{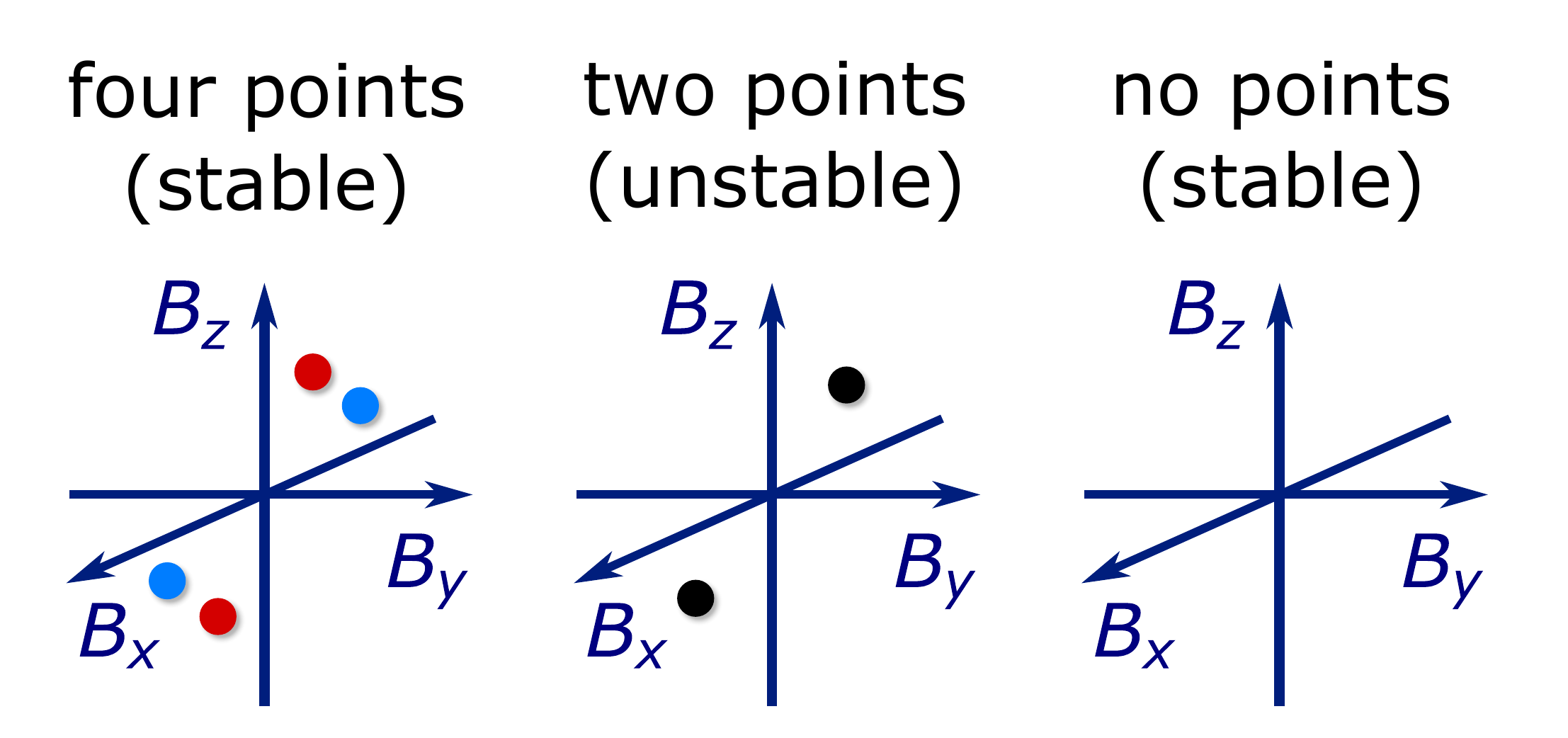}
	\end{center}
	\caption{\textbf{Magnetic degeneracy points for two interacting spin-1/2 electrons, when the relative sign of the $g$-tensors is negative,
	${\cal S}=-1$.}
	Color of each point indicates its topological charge:
	$+1$ (red), $-1$ (blue), 0 (black).
	Four points and no points patterns are stable, 
	and the generic transition between them is that
	the two oppositely charged pairs meet and their charges
	annihilate each other (two points).}
	\label{fig:negativesign}
\end{figure}

In this case, we find two stable eigenpatterns with codimension $d=0$, 
eigenpatterns (VIII) and (XI). 
Pattern (XI) corresponds to the trivial case, with no magnetic degeneracy 
at any field, while pattern  (VIII) to the case of having two 
positively charged and two negatively charged 
Weyl points. 
One of the important messages of Table~\ref{tab:table2} is that there is no 
other stable 
configuration.  The almost stable eigenpattern (X) of two chargeless degeneracy points is the 
generic pattern, separating  extended regions (VIII) and (XI) of stable configurations in parameter space. 
These are precisely the points at which two oppositely charged Weyl points merge and annihilate each other, see Fig.~\ref{fig:negativesign}.
The remaining eigenpattern (IX) corresponds to a loop of magnetic degeneracy points
and, with codimension $d=3$, is very unstable.

\section{Conclusions}
\label{sec:conclusions}

In conclusion, we have provided a full analytical description
of the geometrical patterns and topological charge distribution 
patterns
of magnetic ground-state degeneracy points in a spin-orbit-coupled
two-spin system.
By recognizing the special structure of the Hamiltonian, 
we have mapped  the problem of finding the denegeracy points 
to the eigenproblem of a real non-symmetric $3 \times 3$ 
matrix. 
We have found three drastically different regions 
in parameter space, according to the total topological charge in 
magnetic space, ${\cal Q} = 
\text{sgn}(\text{det}\,\hat{\vec{g}}_{\text{L}}) +  
\text{sgn}(\text{det}\,\hat{\vec{g}}_{\text{R}}) $.

The regions ${\cal Q} = \pm 2$ are very similar, only the signs of the charge 
 distributions are the opposite.  In both regions, 
our stability analysis reveals the existence of two stable and extended (i.e., non-zero measure) regions in the secondary parameter space: in the first one, case (VII), two magnetic Weyl points carry the topological charge, while in the second one, case (I), six Weyl points carry the total charge, 
${\cal Q} = 4\times (+1) + 2\times(-1)$. 
 There is \emph{no other} geometrically stable region, meaning that other charge patterns exist only in special, fine-tuned Hamiltonians, realized for 
 secondary parameters forming a set of zero measure. 
 Some of these configurations can, however, be observed.
For ${\cal Q} = +2$, for instance, the most stable unstable structure, the one 
with two merged Weyl points, that is, when $\text{ep}(\hat{\vec{M}})=$ (III),  
emerges at the \emph{boundary} between the two topologically stable 
phases, (I) and (VII). 
We naturally cross this surface in case we change some of the secondary parameters of the Hamiltonian, 
 such that we go continuously from a region with $\text{ep}(\hat{\vec{M}})=$(I) to a region with $\text{ep}(\hat{\vec{M}})=$(VII).
Approaching this surface, a positively and negatively charged pair of Weyl points must approach each other, and 
just merge to annihilate each other.
This boundary, corresponding to   
$\text{ep}(\hat{\vec M})=$(III),  includes further special patterns, corresponding to further fine-tuning of the parameters.

A similar picture emerges for the case ${\cal Q} = 0$. 
There, two generic, extended regions are found 
with no degeneracy points ($\text{ep}(\hat{\vec M})=$(XI)), and with four
Weyl points ($\text{ep}(\hat{\vec M})=$(VIII)), respectively. 
The generic surface between them corresponds to 
$\text{ep}(\hat{\vec M})=$(X), with two neutral 
degeneracy points, where the two pairs of Weyl points just merged.
These magnetic degeneracy patterns are shown 
in Fig.~\ref{fig:negativesign}.

The linear stability of regions (I), (VII), (VIII) and (XI) 
is corroborated by robust 
topological arguments: 
Weyl points cannot disappear upon infinitesimal smooth deformations
of the secondary parameters.
Starting from these configurations, 
removing Weyl points or creating new degeneracy points
requires fine-tuning of the 
parameters: either positive and negative charges must move towards each 
other, and then annihilate at a very special point (this corresponds to the 
boundaries (III) and (X) discussed above), 
or fine-tuning leads to the formation of an ellipse or an ellipsoid of
degeneracy points that are not Weyl points, as their energy dispersion
is flat in at least one direction. 
Numbers and charges of the Weyl points 
in regions (I), (VII), (VIII) and (XI) are constrained by  
the total topological charge,
whereas the relative locations of these Weyl points is constrained
by the fact that they always  
come in time-reversed pairs.\cite{Scherubl}

The classification problem studied here is readily generalized
to any interacting spin system, e.g., three interacting electrons subject
to spin-orbit coupling, or  two interacting spins with larger spin size.
Studying possible geometrical patterns of the
magnetic degeneracy points, the corresponding
topological charge density patterns, the stability 
of these and their evolution upon continuous deformations of the
Hamiltonian are interesting and challenging tasks. 

A further direction for generalization is to analyze
magnetic degeneracy points of higher-energy
eigenstates instead of the ground state\cite{Bruno}.
An interesting set of further problems is obtained if the 
spin-orbit effects are not completely arbitrary, but
are subject to symmetry constraints. For example, if magnetic adatoms are placed on specific sites on 
a metallic surface\cite{Wiesendanger,Spinelli}, then the 
spatial symmetry of the arrangement will restrict the 
form of the $g$-tensors and the exchange terms. 
In general, a fully numerical approach can provide insight
into the questions above, but it seems difficult to efficiently
generalize the analytical treatment used here
to obtain exact results for a much broader set of physically 
motivated Hamiltonians.

Besides opening up interesting theory questions, we hope that
our work will inspire experiments as well. 
Weyl points in interacting spin systems have been found
experimentally by transport\cite{Scherubl} and Landau-Zener
spectroscopy\cite{Wernsdorfer}, but their topological characterization, e.g., 
via measurements of the Berry curvature\cite{Gritsev,Roushan}, 
is yet to be done.
To observe a sharp transition between different eigenpatters studied here,
e.g., to observe creation and annihilation of Weyl points, 
the experimenter needs exquisite control over the $g$-tensors and
spin-spin interaction. 
This is given, to some extent, in spin-orbit-coupled quantum dots\cite{Csonka,VeldhorstPRB},
but a fruitful alternative realization could be done 
in superconducting qubits, where synthetic qubit-qubit interaction, 
including antisymmetric exchange\cite{DaWeiWang}, can be engineered and controlled.



\acknowledgments
We thank P. Vrana for useful discussions. 
This work was supported by the National Research Development and Innovation Office of Hungary within the Quantum Technology National Excellence Program (Project No. 2017-1.2.1-NKP-2017-00001), under OTKA Grant 124723, 127900, by the New National Excellence Program of the Ministry of Human Capacities,
by the QuantERA SuperTop project, and by the AndQC FetOpen project.

\appendix

\section{Survival kit for the eigenvalue problem of non-symmetric real 
matrices}
\label{app:linalg}

Quantum physicists are very familiar with the eigenvalue problem of
symmetric real matrices and Hermitian complex matrices, but less
familiar with the eigenproblem of real non-symmetric matrices. 
Therefore, here we summarize a few useful concepts
and relations about the latter.

For concreteness, we consider a $3 \times 3$ real 
non-symmetric matrix $\hat{\vec{M}}$, as in the main text. 
The complex number $\lambda$ is an eigenvector of $\hat{\vec{M}}$, 
if $\text{det}(\hat{\vec M} - \lambda \mathds{1}) = 0$.
All eigenvalues of a real symmetric matrix are real, but
this is not guaranteed if the matrix is non-symmetric. 
The quantity $\text{det}(\hat{\vec M} - \lambda \mathds{1})$
is called the characteristic polynomial of $\hat{\vec{M}}$. 
According to the fundamental theorem of algebra, the 
characteristic polynomial can be factored into the 
product of 3 terms, 
\bean
\text{det}(\hat{\vec M} - \lambda \mathds{1}) =
(\lambda_1 - \lambda)(\lambda_2-\lambda)(\lambda_3 - \lambda),	
\eean
where $\lambda_i$-s are the eigenvalues of $\hat{\vec{M}}$. 

The algebraic multiplicity $\mu(\lambda_i)$ of the eigenvalue
$\lambda_i$ is its multiplicity as a root of the characteristic
polynomial. 
The geometric multiplicity $\gamma(\lambda_i)$ is
the maximal number of linearly independent eigenvectors
belonging to the eigenvalue $\lambda_i$, that is, 
the dimension of the eigensubspace corresponding to $\lambda_i$. 
For generic $3\times 3$ matrices, 
it holds that $1 \leq \gamma(\lambda_i) \leq \mu(\lambda_i) \leq 3$, 
while for real symmetric and complex Hermitian matrices, 
it holds that $1 \leq \gamma(\lambda_i) = \mu(\lambda_i) \leq 3$.

A real symmetric (complex Hermitian) matrix can always be
diagonalized with an orthogonal (unitary) transformation, 
whose matrix is constructed from the eigenvectors. 
This is not the case for a real non-symmetric matrix. 
Nevertheless, there is still a canonical form $\hat{\vec J}$ of a 
real non-symmetric 
matrix $\hat{\vec M}$, called the Jordan
normal form, see the examples in 
Table~\ref{tab:table1}.
The Jordan normal form is not diagonal, 
but it has a block-diagonal structure, 
and this structure is related to the
algebraic and geometrical multiplicities of $\hat{\vec M}$.
If the matrix has a size greater than 3, then the
eigenvalues and their algebraic and 
geometric multiplicities are in general 
not sufficient to determine the Jordan normal form of the matrix. 
But in our case, when $\hat{\vec M}$ is a $3 \times 3$ matrix, 
the eigenvalues and the multiplicities do determine the
Jordan normal form: each eigenvalue $\lambda_i$ 
has a block in the Jordan normal
form with the size of its algebraic multiplicity $\mu(\lambda_i)$, 
and within each block, there are as many 1 entries on the superdiagonal 
as the difference $\mu(\lambda_i) - \gamma(\lambda_i)$
between the algebraic and geometric multiplicities. 
In Table~\ref{tab:table1}, we list 7 examples. 
The claim is that for any matrix $\hat{\vec M}$, there is a similarity 
transformation (invertible matrix) $\hat{\vec P}$, such that 
\bean
\hat{\vec J} = \hat{\vec P}^{-1} \hat{\vec M} \hat{\vec P}
\eean
is a Jordan normal form, and hence any matrix $\hat{\vec M}$ can
be decomposed as in Eq.~\eqref{eq:jordandecomposition}.

\section{The role of local spin basis transformations}
\label{app:Minv}

Here, we collect a few facts and remarks about the role of local spin basis 
transformations in the spin-orbit-coupled two-spin problem 
studied in the main text. 
Some of these are used in our proofs. 

\emph{Existence of a basis where the $g$-tensors are symmetric.}
The Hamiltonian in Eq.~\eqref{eq:hamiltonian} is built on a 
single-particle model\cite{Scherubl} in which both sites 
support a single spinful orbital level, and the corresponding two states
are Kramers degenerate at zero magnetic field. 
Because of this Kramers degeneracy, there is an ambiguity in choosing
the basis states.
The Hamiltonian in Eq.~\eqref{eq:hamiltonian} is generic, i.e., 
this form is guaranteed no matter how we choose the basis. 
However, the actual secondary parameters ($g$-tensor matrix elements
and the rotation matrix $\hat{\vec R}$) change if we change the basis.

First, we show that one can always choose 
a local spin basis in which the $g$-tensors are symmetric. 
For this, we recall that any local spin basis transformation,
apart from an arbitrary complex phase, can be written
as a unitary operation\cite{Diosi}
\bean
U(\vec \alpha) = \exp\left(-i \vec \alpha \vec{S}\right),
\eean
where $\vec \alpha$ is a real three-component vector.
Furthermore, this transformation corresponds to a rotation
$\hat{\vec O}_{\vec \alpha}$
of the spin vector operator around the direction of $\vec \alpha$
with angle $\alpha = |\vec \alpha|$:
\bean
U(\vec \alpha) \vec S U(\vec \alpha)^\dag
= \hat{\vec O}_{\vec \alpha} \vec S.
\eean

When we apply two different transformations on the two sites, 
$U_\text{L}$ and $U_\text{R}$, 
which are represented by the rotations 
$\hat{\vec O}_\text{L}$ and $\hat{\vec O}_\text{R}$, 
respectively, 
then
this combined transformation results in 
\bean
H' &=& (U_\text{L} \otimes U_\text{R})
H (U_\text{L} \otimes U_\text{R})^\dag
\nonumber \\
&=&
\mu_{\text{B}}\vec B(\hat{\vec g}_{\text{L}}\hat{\vec O}_{\text{L}}\vec S_{\text{L}}+\hat{\vec g}_{\text{R}}\hat{\vec O}_{\text{R}}\vec S_{\text{R}})+\nonumber \\
&+&
J \vec S_{\text{L}} \hat{\vec O}_{\text{L}}^{-1}\hat{\vec R}
\hat{\vec O}_{\text{R}}\vec S_{\text{R}}.
\label{eq:spintransformed}
\eean

Recall also that any real symmetric matrix has a polar decomposition
to a symmetric real matrix and a rotation matrix, 
therefore the $g$-tensors can also be decomposed
as $\hat{\vec g}_\text{L/R} = \hat{\vec G}_{\text{L/R}} 
\hat{\vec Q}_\text{L/R}$,
where $\hat{\vec G}_\text{L/R}$ are real and symmetric and
$\hat{\vec Q}_\text{L/R}$ are rotations.
Therefore, by using the basis transformation 
as $\hat{\vec O}_\text{L/R} = \hat{\vec Q}_\text{L/R}^{-1}$, 
the transformed Hamiltonian reads
\bean
H' &=&
\mu_{\text{B}}\vec B(\hat{\vec G}_{\text{L}} \vec S_{\text{L}}
+
\hat{\vec G}_{\text{R}}
\vec S_{\text{R}})+\nonumber \\
&+&
J \vec S_{\text{L}} \hat{\vec R}' \vec S_{\text{R}}.
\eean
 
An interesting feature of this particular basis choice is
its strong relevance to spectroscopy experiments:
if a spectroscopy experiment measures the Zeeman splitting
as a function of the magnetic field direction, then 
the principal axes and principal values of $\hat{\vec G}$
can be directly calculated from that data.\cite{Crippa} 
Another interesting feature of this basis 
is that in the limit of small exchange rotation, $\hat{\vec R}\rightarrow\mathds{1}$,
the degeneracy points always form the `six-point' pattern,
due to the fact that the matrix $\hat{\vec M}$ converges
to a real matrix that has only positive eigenvalues.

\emph{$\hat{\vec M}$ is invariant under any
local spin transformation.}
In the main text, we have used the matrix $\vec{\hat M}$ 
to characterize the magnetic degeneracy points. 
It is a natural expectation that the locations of the
degeneracy points do not depend on the local spin basis choice. 
Can we prove this directly? 
Yes, and actually we can prove something stronger:
the matrix $\vec{\hat M}$ itself is invariant 
under local spin basis transformations. 
This is a straightforward consequence
of the transformation rules used above.
A general basis transformation results in 
$\hat{\vec g}_\text{L/R} \mapsto
\hat{\vec g}_{\text{L,R}}\hat{\vec O}_{\text{L,R}}$
and
$\hat{\vec R} \mapsto \hat{\vec O}_{\text{L}}^{-1}\hat{\vec R}\hat{\vec O}_{\text{R}}$. 
Substituting these into the definition of $\hat{\vec M}$ we can get the transformed matrix as
\bean
\hat{\vec M} \mapsto 
(\hat{\vec g}_{\text{L}}\hat{\vec O}_{\text{L}})(\hat{\vec O}_{\text{L}}^{-1}\hat{\vec R}\hat{\vec O}_{\text{R}})(\hat{\vec O}_{\text{R}}^{-1}\hat{\vec g}_{\text{R}}^{-1})=\hat{\vec M},
\eean
so $\hat{\vec M}$ is indeed invariant under local spin transformations as we expected.

\section{Locations of the magnetic degeneracy points}
\label{app:locations}

Here, we outline how to determine the locations of the 
ground-state magnetic degeneracy points studied in the
main text.

First, we consider the case when 
the magnetic field vector $\vec{B}=B\vec{v}_a$ is 
a left eigenvector of 
$\hat{\vec M}$ with a positive eigenvalue $a$, and we
assume that $\vec{v}_a$ is normalized.
Then, the effective magnetic field vectors of Eq.~\eqref{eq:transformed}
are parallel, $\vec{B}_\text{L,eff} \parallel \vec{B}_\text{R,eff}$,
and point to the same direction. 
Using that direction as the spin quantization axis, 
the Hamiltonian defined in Eq.~\eqref{eq:transformed} has a special form 
\bean\label{eq:twospiniso}
H' =\mu_{\text B} B (g_{\text{L}}  S'_{\text{L}z}+g_{\text{R}}S'_{\text{R}z})+J\vec S'_{\text{L}}\cdot\vec S'_{\text{R}},
\eean
where $g_{\text{L,R}}=|\vec v_{a}\hat{\vec g}_{\text{L,R}}|$. 
Later we will use the fact
that 
\bean
\label{eq:ratio}
g_{\text{L}}/g_{\text{R}}=a,
\eean
which 
follows from $\vec v_a$ being a left eigenvector of $\hat{\vec M}$
with eigenvalue $a$. 
This Hamiltonian $H'$ 
conserves the total 
spin projection 
$S_{z}'=S_{\text{L}z}'+S_{\text{R}z}'$,
and thus has the following block-diagonal matrix form in the product basis
$\ket{\uparrow \uparrow}$, 
$\ket{\uparrow \downarrow}$, 
$\ket{\downarrow \uparrow}$, 
$\ket{\downarrow \downarrow}$:
\bean
\label{eq:hamiltonmatrix}
H' =\frac{1}{2} 
\begin{pmatrix}
\mu_{\text{B}}g_{+}B&0&0&0\\
0&\mu_{\text{B}}g_{-}B-J&J&0\\
0&J&-\mu_{\text{B}}g_{-}B-J&0\\
0&0&0&-\mu_{\text{B}}g_{+}B\\ 
\end{pmatrix},
\eean
where 
$g_{\pm}=g_{\text{L}}\pm g_{\text{R}}$. Energy eigenstates  from different 
subspaces of $S_{z}'$ can be degenerate, because there are no matrix 
elements mixing them.

At zero magnetic field, the ground state of
$H'$ in Eq.~\eqref{eq:hamiltonmatrix} 
is the singlet state 
$(\ket{\uparrow \downarrow} - \ket{\downarrow \uparrow})/\sqrt{2}$
from the $S_{z}'=0$ subspace, with energy $-J$ , 
and the remaining three states are triplets with zero energy.
If the magnetic field is much greater than the interaction strength $J$, 
then the energy eigenstates are the product states. 
The ground state is the state $\ket{\downarrow \downarrow}$ from the 
$S_{z}'=-1$ subspace with energy $-\frac{1}{2}\mu_{\text{B}}g_{+}B$:
this follows from that fact that $|g_+| > |g_-|$, which is implied by
$g_\text{L}, g_\text{R} > 0$.
Therefore, at a certain magnetic field strength between zero and 
infinity, the $S_{z}'=0$ ground state must be degenerate with the
$S_{z}'=-1$ ground state. 
In fact, straightforward calculation shows that this 
level crossing degeneracy happens
at the critical magnetic field strength
\bean\label{eq:isobfield}
B_{a}=\frac{J}{2\mu_B}\left(\frac{1}{g_{\text{L}}}+\frac{1}{g_{\text{R}}}\right),
\eean
and the degenerate ground states are
\begin{subequations}
\bean
\ket{0}&=&\frac{1}{\sqrt{g_{\text L}^{2}+g_{\text R}^{2}}}(g_{\text R}\ket{\uparrow\downarrow}-g_{\text L}\ket{\downarrow\uparrow})\\
\ket{-1}&=&\ket{\downarrow\downarrow}
\eean
\label{eq:states}
\end{subequations}
labelled with their $S'_z$ quantum number.

If, on the other hand, $\vec B$ is a
left eigenvector of $\hat{\vec M}$ with a negative eigenvalue, 
then the effective magnetic fields
$\vec{B}_\text{L,eff}$ and $\vec{B}_\text{R,eff}$
 are anti-aligned.
Then, the Hamiltonian can be brought to the same form as in 
Eq.~\eqref{eq:hamiltonmatrix}, with the change
that now $|g_+| < |g_-| $.
Therefore, the ground states in the limits
of zero and large magnetic fields are both 
in the $S_{z}'=0$ subspace, 
and there is no ground-state level crossing.

\section{Closed degeneracy lines are ellipses, closed 
degeneracy surfaces are ellipsoids}
\label{app:ellipse}

In Table \ref{tab:table1}, the eigenpattern (IV) implies 
that the degeneracy points form a closed surface. 
Here we show that this surface is an ellipsoid. 
A similar proof shows that the loops formed by the degeneracy 
points in cases (II), (V), and (X) are ellipses.

In case (IV), the matrix $\hat{\vec M}$ has a single eigenvalue $a$
and the normalized eigenvectors form the three-dimensional unit sphere. 
So, any $\vec v$ unit vector is an eigenvector, and
we can apply the results \eqref{eq:ratio} and \eqref{eq:isobfield}
to obtain the locations of the degeneracy points
\bean
\vec B_{a}(\vec v)=\frac{J(1+a)}{2\mu_{\text{B}}}\frac{\vec v}{|\vec v\hat{\vec g}_{\text{L}}|} 
=
\frac{J(1+a)}{2\mu_{\text{B}}}\frac{\vec v}{|\vec v\hat{\vec G}_{\text{L}}|}
.
\eean
In the second step, we have made use of the 
polar decomposition
of $\hat{\vec g}_\text{L}$, introduced in Appendix \ref{app:Minv},
where $\hat{\vec G}_\text{L}$ denotes the real symmetric
component. 
In the principal reference frame of $\hat{\vec G}_\text{L}$, 
the location of the degeneracy point 
associated to $\vec v$ 
reads 
\bean
\left(\bna{c}
B_{ax} \\
B_{ay} \\
B_{az} 
\eda \right) 
=\frac{J(1+a)}{2\mu_{\text{B}}}\frac{1}{\sqrt{G_{x}^{2}v_{x}^{2}+G_{y}^{2}v_{y}^{2}+G_{z}^{2}v_{z}^{2}}}
\left( \bna{c}
v_{x} \\
v_{y} \\
v_{z}
\eda \right),
\eean
where $(G_x,G_y,G_z)$ are the principal values
of $\hat{\vec G}_\text{L}$.
Acting with $\hat{\vec G}_\text{L}$ on both sides of the
equation, and taking the length-squared of the resulting vectors, 
we obtain the equation 
\bean
G_{x}^{2}B_{ax}^{2}+G_{y}^{2}B_{ay}^{2}+G_{z}^{2}B_{az}^{2}=\left[\frac{J(1+a)}{2\mu_{\text{B}}}\right]^{2},
\eean
which implies that the degeneracy points 
form an ellipsoid. 

For cases (II), (V), and (X),
we have an additional constraint: $\vec v$ has to be 
in the degenerate subspace of $\hat{\vec M}$. 
This intersects the ellipsoid 
with a plane passing through the origin.
Since the intersection of a plane and an ellipsoid is always
an ellipse, we conclude  that 
the degeneracy points in these cases are ellipses.

\section{Topological charge distributions}
\label{app:charge}

Here, we outline the derivation of the topological charges associated
to the ground-state magnetic degeneracy points.
The results were summarized in Tables \ref{tab:table1} and \ref{tab:table2}
in the main text.  
The first, simple step of the derivation is to approximate
the Hamiltonian in the vicinity of the degeneracy point and 
truncate it to the two-dimensional degerate subspace of interest.
A second, nontrivial step is to connect this two-dimensional
Hamiltonian to the eigenvalue problem of the matrix $\hat{\vec M}$,
which allows us to express the topological charges
of the degeneracy points via the eigenvalues of $\hat{\vec M}$.

To exemplify the derivation,  
consider the case when the total topological charge is 
${\cal Q} = +2$, and 
the eigenpattern of $\hat{\vec M}$ is
(I) (see Table~\ref{tab:table1}).
Then $\hat{\vec M}$ has
three eigenvalues $a$, $b$, $c$, 
three eigenvectors $\vec v_a$, $\vec v_b$, $\vec v_c$,
and the set of the ground-state magnetic degeneracy points
is formed by six Weyl points.
To calculate the topological charge of a Weyl point, 
say, $\vec B_{a+}$, we focus on the two degenerate ground
states $\ket{0}$ and $\ket{-1}$ in the degeneracy point
(see Eqs.~\eqref{eq:states}), 
make a linear expansion of the Hamiltonian for 
small deviations $\delta \vec B = \vec B - \vec B_{a+}$ 
of the magnetic field from the degeneracy point, 
and truncate the Hamiltonian for the two-dimensional 
subspace spanned by $\ket{0}$ and $\ket{-1}$. This reduced Hamiltonian can be written in terms of Pauli matrices,
\bean
\label{eq:effectivegtensor}
H_\text{red}(\delta \vec B) = \mu_{\text{B}}\, \delta \! \vec B\,  \hat{\vec g}_{a+} \vec \tau,
\eean
where $\vec \tau = (\tau_1,\tau_2,\tau_3)$ is half times the vector of Pauli
matrices, e.g., $\tau_3 = \frac{1}{2}(\ket{0}\bra{0} - \ket{-1}\bra{-1})$.
Because of the similarity of $H_\text{red}$ and the Hamiltonian 
of a spin in a magnetic field with an anisotropic $g$-tensor, 
we call $\hat{\vec g}_{a+}$ the \emph{effective $g$-tensor}
of the degeneracy point $\vec{B}_{a+}$.
The determinant of effective $g$-tensor of a Weyl point is nonzero, 
and its sign provides the topological charge of the Weyl 
point:
\bean
\mathcal{Q}_{a+} = \text{sgn} \, \text{det} \left(\hat{\vec g}_{a+}\right).
\eean 

To obtain an analytical result for the elements of the effective $g$-tensor, 
we evaluate $H_\text{red}$ in Eq.~\eqref{eq:effectivegtensor}
with \mbox{$\delta \vec{B} =\delta B \vec{e}_\alpha$ ($\alpha = x, y, z$)}
pointing along the unit vector $\vec{e}_\alpha$ of direction $\alpha$,
multiply both sides with \mbox{$\tau_\beta$ ($\beta = x, y, z$)},
and take the trace of both sides. 
This procedure yields the matrix elements 
\bean
\label{eq:charge}
(\hat{\vec g}_{a+})_{\alpha\beta}=
\frac{2}{\mu_{\text{B}}\delta B}\text{Tr}[H_{\text{red}}(\delta B\,  \vec e_{\alpha})\cdot\tau_{\beta}].
\eean
The matrix obtained from this relation can be identified with
this expression:
\bean
\begin{split}\label{eq:effectivegtensor2}
\hat{\vec g}_{a+}=&\frac{g^{2}_{\text{R}}\hat{\vec g}_{\text{L}}+g^{2}_{\text{L}}\hat{\vec g}_{\text{R}}\hat{\vec R}^{-1}}{g^{2}_{\text{L}}+g^{2}_{\text{R}}}\vec e_{z}\otimes\vec e_{z}+\\
&\frac{g_{\text{R}}\hat{\vec g}_{\text{L}}-g_{\text{L}}\hat{\vec g}_{\text{R}}\hat{\vec R}^{-1}}{\sqrt{g^{2}_{\text{L}}+g^{2}_{\text{R}}}}(\mathds{1}-\vec e_{z}\otimes\vec e_{z}).
\end{split}
\eean
Here, $\otimes$ denotes the dyadic product.

The determinant of \eqref{eq:effectivegtensor2} can 
expressed with the eigenvalues of the matrix $\hat{\vec M}$, yielding
\begin{equation}\label{eq:effgdet}
\det\hat{\vec g}_{a+}=\frac{a(1+a)}{(1+a^{2})^{2}}(a-b)(a-c) \det\hat{\vec g}_{\text{R}},
\end{equation}
Inserting this determinant into Eq.~\eqref{eq:charge} and
using $a>0$ and $\text{det} \, \hat{\vec g}_\text{R} > 0$, 
we obtain 
\begin{equation}
{\cal Q}_{a+}=\text{sgn}[(a-b)(a-c)].
\end{equation}
The same result is obtained for ${\cal Q}_{a-}$, and analogous results
are obtained for the remaining four degeneracy points, e.g., 
\begin{equation}
{\cal Q}_{b+}=\text{sgn}[(b-a)(b-c)].
\end{equation}
These results imply that for the eigenpattern (I), 
the distribution of topological charge among the 
six degeneracy points is $4\times(+1),\ 2\times(-1)$, and 
that the negatively charged point pair belongs to the eigenvalue
that is between the other two eigenvalues. 

For the eigenpatterns (II) and (III), the two Weyl points 
belonging to the non-degenerate eigenvalue $b$ can be 
analyzed in an analogous fashion, with the result that
their topological charge is $+1$. 
As a natural consequence of this and the sum rule that the
total topological charge is +2, the remaining degeneracy points,
that is, the ellipse in case (II) and the two remaining points in case (III), 
must have zero topological charge. 
For the remaining eigenpatterns from (III) to (VII), 
the distribution of the total topological charge $+2$ is obvious. 

We note that the result \eqref{eq:effgdet} is
valid not only for Weyl points but for any
ground-state magnetic degeneracy point.
This has interesting implications regarding the energy dispersion
in the vicinity
of a degeneracy point $\vec B_{a+}$ whenever that point
is in an eigenspace of $\hat{\vec M}$ that
belongs to a degenerate eigenvalue $a$.
In that case, the degeneracy of $a$ implies that the 
right hand side of \eqref{eq:effgdet} yields zero, i.e., there is 
at least one direction for $\delta \vec{B}$ along which the dispersion
is non-linear. 
In cases (II), (IV), (V) and (IX), naturally, 
the special non-dispersive directions
are the tangents of the ellipse and the ellipsoid. 
However, it is remarkable that discrete degeneracy points can 
also have non-linear dispersion.
Examples are the two points in case (VI) with non-zero charge and the 
neutral points in cases (III) and (X). 
Degeneracy points showing similar non-linear dispersion
are sometimes called multi-Weyl points\cite{ChenFang_multiweyl,ZhongboYan,Ahn,ZeMinHuang} 
in the literature.
In general, their topological charges cannot be determined by
their effective $g$-tensor: for example, in case (VI), 
the determinants of the effective $g$-tensors of the two 
degeneracy points are zero, nevertheless each point
has a topological charge $+1$.

\bibliography{weyl-NoCorrections}

\end{document}